\documentclass[prl,superscriptaddress,nofootinbib,showpacs,twocolumn]{revtex4-2}

\usepackage[english]{babel}

\usepackage[utf8]{inputenc}
\usepackage{amsmath,amssymb}
\usepackage{bbold}
\usepackage{slashed}
\usepackage{subfigure}
\usepackage[colorlinks=true,
breaklinks=true,
urlcolor=blue,
citecolor=blue]{hyperref}
\usepackage[usenames,dvipsnames]{color}
\usepackage{ulem}
\usepackage{appendix}   
\usepackage{braket,bm}
\usepackage{multirow}
\usepackage{enumitem}
\usepackage{color}
\usepackage{cancel}
\usepackage{dsfont}
\usepackage{orcidlink}
\usepackage{array}
\usepackage{indentfirst}
\usepackage{bbm}
\usepackage{epstopdf}
\usepackage{graphicx}
\usepackage{caption}
\DeclareCaptionLabelFormat{figprefix}{FIG.#2.} 
\allowdisplaybreaks

\AtBeginDocument{}

\begin{document}

\title{Toward a holographic realization of the 2+1-flavor QCD phase structure}

\author{Jin-Yang Shen}

\affiliation{School of Physics and Electronics, Hunan University, Changsha 410082, China}

\author{Xin-Yi Liu}

\affiliation{School of Fundamental Physics and Mathematical Sciences, Hangzhou Institute for
 Advanced Study, UCAS, Hangzhou 310024, China}
 
\affiliation{Institute of Theoretical Physics, Chinese Academy of Sciences, Beijing 100190, China}

\affiliation{University of Chinese Academy of Sciences (UCAS), Beijing 100049, China}

\author{Jin-Rui Wu}

\affiliation{School of Physics and Electronics, Hunan University, Changsha 410082, China}

\author{Yue-Liang Wu}
\affiliation{School of Fundamental Physics and Mathematical Sciences, Hangzhou Institute for Advanced Study, UCAS, Hangzhou 310024, China}
\affiliation{Institute of Theoretical Physics, Chinese Academy of Sciences, Beijing 100190, China}
\affiliation{University of Chinese Academy of Sciences (UCAS), Beijing 100049, China}
\affiliation{International Center for Theoretical Physics Asia-Pacific (ICTP-AP), UCAS, Beijing 100190, China}

\author{Zhen Fang}
\email{zhenfang@hnu.edu.cn}
\affiliation{School of Physics and Electronics, Hunan University, Changsha 410082, China}

\affiliation{Hunan Provincial Key Laboratory of High-Energy Scale Physics and Applications, Hunan University, Changsha 410082, China}


\begin{abstract}
We present a fully back-reacted Einstein--Maxwell--Dilaton--flavor model with dynamical light and strange sectors, calibrated to lattice QCD using a machine-learning--assisted spectral method. The model reproduces the 2+1-flavor equation of state and chiral dynamics with quantitative accuracy, and maps the Columbia plot with a tri-critical point at \(m_s^{\mathrm{tri}} \simeq 21~\text{MeV}\) and a critical mass \(m_c \simeq 0.785~\text{MeV}\), consistent with lattice results. At finite density, it yields a crossover-to-first-order transition and predicts a critical endpoint at \(T_C = 75.4~\text{MeV}\) and \(\mu_C = 768~\text{MeV}\), within the reach of heavy-ion experiments. These findings establish a unified holographic framework for the QCD phase structure across quark masses and baryon density, providing the first consistent and quantitative description of both deconfinement and chiral transitions within a single holographic model.
\end{abstract}

\maketitle

\newpage


\textit{Introduction}. The phase structure of Quantum Chromodynamics (QCD) lies at the core of our understanding of strongly interacting matter. It dictates how deconfinement and chiral symmetry restoration emerge and evolve across a wide range of temperatures, baryon chemical potentials, and quark masses, which are crucial for interpreting the early universe, the interiors of neutron stars, and the hot, dense medium created in heavy-ion collisions \cite{Aoki:2006we,Fukushima:2025ujk}. Decoding this structure, both qualitatively and quantitatively, is vital for understanding the nature of confinement, chiral symmetry breaking, and their interplay in the QCD medium, which remains a central goal in theoretical nuclear physics.

A central aspect of the QCD phase structure at zero chemical potential is its dependence on the light and strange quark masses, as illustrated by the Columbia plot \cite{Brown:1990ev, Laermann:2003cv}. This diagram encodes how the nature of the QCD transition varies across the \((m_{u,d}, m_s)\) plane, reflecting the underlying chiral and center symmetries of QCD. In the two-flavor chiral limit \((m_{u,d} \to 0)\), a second-order phase transition is expected, possibly in the O(4) universality class depending on the strength of the axial anomaly \cite{Pisarski:1983ms}; whereas for three massless flavors, the transition is generally first-order. A tri-critical point is believed to exist along the \(m_{u,d} = 0\) boundary at a critical strange quark mass \(m_s^{\text{tri}}\), separating these regimes, although its precise location remains uncertain. The Columbia plot thus provides not only a symmetry-based classification of QCD transition behavior, but also a stringent test for any theoretical framework aiming to reproduce the full QCD phase structure.

While lattice QCD has revealed key features of the QCD phase diagram at zero chemical potential \cite{Borsanyi:2013bia,HotQCD:2014kol}, a complete and continuous description of QCD thermodynamics and chiral dynamics across varying quark masses and finite baryon densities remains out of reach. In this context, gauge/gravity duality provides a powerful non-perturbative framework for modeling strongly coupled gauge theories \cite{Witten:1998qj, Gubser:1998bc,Maldacena:1997re}. A long-standing challenge in holographic QCD, however, lies in constructing a unified and dynamically consistent model that faithfully reproduces the full QCD phase structure, including both chiral and deconfinement transitions at zero and finite chemical potential \cite{Gubser:2008yx,DeWolfe:2010he,Jarvinen:2011qe,Chelabi:2015cwn,Li:2016smq,Arefeva:2022avn,Critelli:2017oub,Li:2012ay,Fang:2019lsz,PhysRevD.103.086021,Chen:2019rez,Alho:2012mh,Alfano:2025dch,Li:2014hja,Alho:2013hsa,Fu:2024nmw,Cai:2012xh,Cai:2024eqa,Zhang:2025vsm,Grefa:2022sav,Zhao:2022uxc,Li:2025lmp,Demircik:2024aig,Ecker:2025vnb}. In particular, capturing the quark mass dependence encoded in the Columbia plot remains a key benchmark for the success of any such approach.

In this work, we develop a fully back-reacted Einstein-Maxwell-Dilaton-flavor (EMDf) model that incorporates dynamical light and strange quark sectors through independent scalar fields, consistently coupled to the gravity-dilaton-Maxwell background. To address challenges in high-dimensional parameter space, we combine machine learning with spectral methods to calibrate the model, reproducing lattice QCD results for the equation of state (EoS) and chiral condensates at zero chemical potential. The resulting framework provides a quantitatively accurate holographic realization of the 2+1-flavor QCD phase structure. It successfully reproduces the Columbia plot, capturing both the flavor-symmetric critical mass and the boundary separating the first-order and crossover regions. Furthermore, the model naturally extends to finite baryon chemical potential, yielding a consistent phase diagram in the \((T, \mu_B)\) plane, including the emergence of a critical endpoint (CEP).

Our results demonstrate, for the first time, that the full QCD phase structure---spanning both quark mass dependence and finite chemical potential---can be consistently and accurately realized within a bottom-up holographic framework. This work represents a significant advancement in holographic QCD modeling, and highlights the potential of data-driven, non-perturbative approaches to unravel the rich dynamics of strongly interacting matter.


\textit{The EMDf system with 2+1 flavors}. We consider the \(2+1\)-flavor EMDf model described by the following action:
\begin{align}\label{Stotal}
        S &= \frac{1}{2\kappa_N^2} \int d^5x \sqrt{-g} \left[R -\omega(\phi) F_{ab}F^{ab} - \frac{4}{3}(\partial \phi)^2 - V_E(\phi) \right. \notag\\
         &\quad \left. - \beta e^{\phi} \left((\partial\chi_u)^2 + \frac{1}{2}(\partial\chi_s)^2 + V(\chi_u,\chi_s,\phi)\right) \right],
\end{align}
where \(\kappa_N = \sqrt{8 \pi G_5}\) is the five-dimensional gravitational constant. The Abelian gauge field \(A_a\) introduces the chemical potential, while the dilaton field \(\phi\) encodes the dynamics of non-conformality and confinement \cite{Gubser:2008yx}. The scalar fields \(\chi_u\) and \(\chi_s\) are dual to the boundary chiral operators \(\bar{\psi}\psi\) for the light and strange quarks, respectively. The prefactor \(e^{\phi}\) originates from the soft-wall factor \(e^{-\phi}\) in the string-frame flavor action after the frame transformation \cite{Liu:2023pbt}. The parameter \(\beta\) characterizes the interaction strength between the flavor sector and the bulk geometry. Without loss of generality, we set \(\beta = 1\), since its effects can be absorbed into suitable rescalings of the scalar vacuum expectation values (VEVs) \(\chi_u\) and \(\chi_s\).

This framework can be viewed as an improved 2+1-flavor soft-wall AdS/QCD model consistently coupled to an Einstein-Maxwell-Dilaton background. To reproduce realistic QCD thermodynamics, a nontrivial dilaton potential is introduced in the form:
\begin{equation}
    V_E(\phi) = \frac{1}{L^2} \left(-12 \cosh \gamma_1 \phi_c + b_2 \phi_c^2 + b_4 \phi_c^4\right),
\end{equation}
with a field redefinition $\phi_c = \sqrt{8/3} \, \phi$ ensuring a canonical kinetic term. The coefficient $b_2$ is constrained by the mass-dimension relation, yielding $b_2 = 6\gamma_1^2 - 3/2$.

The Maxwell-dilaton coupling function is chosen as
\begin{equation}
    \omega(\phi) = c_1 \, \text{sech}(c_2 \phi^3 + c_3),
\end{equation}
which ensures a nontrivial dilaton dependence of the gauge dynamics in the bulk.

The flavor potential, encoding the dynamics of the light and strange quark condensates, is defined as:
\begin{equation}
    V(\chi_u, \chi_s, \phi) = e^{\frac{4\phi}{3}} V_\chi(\chi_u, \chi_s, \phi),
\end{equation}
where
\begin{align}\label{Vchichiusphi}
    V_\chi(\chi_u,\chi_s,\phi) &= -\frac{1}{2} \left(3 + \Phi(\phi)\right) (2\chi_u^2 + \chi_s^2) \notag\\
    &\quad + \frac{\gamma}{2\sqrt{2}} \chi_u^2 \chi_s + \frac{\lambda}{4} (2\chi_u^4 + \chi_s^4),
\end{align}
and $\Phi(\phi) = d_1 \phi + d_2 \phi^2$ introduces dilaton-dependent corrections to the effective mass terms \cite{Liu:2023pbt}. The cubic term $\chi_u^2 \chi_s$ originates from the ’t~Hooft determinant $\det[X]$ structure of the flavor sector \cite{Chelabi:2015cwn,Fang:2018vkp}.

We adopt the following metric ansatz for a black hole in asymptotically AdS$_5$ spacetime:
\begin{equation}\label{metric1}
    ds^2 = \frac{L^2 e^{2 A_E(z)}}{z^2} \left(-f(z)dt^2 + \frac{dz^2}{f(z)} + dx^i dx^i\right),
\end{equation}
where $i = 1,2,3$ and $L$ is the AdS radius (set to $L = 1$ hereafter). The holographic coordinate $z$ spans from the boundary ($z \to 0$) to the black hole horizon ($z = z_h$), and $A_E(z)$ characterizes deviations from the pure AdS geometry.

The equations of motion derived from the action \eqref{Stotal} are:
 \begin{align}
         f'' - \frac{3f'}{z} + 3A_E'f' - 4 z^2 e^{-2A_E}\omega(\phi) A_t'^2 &= 0,  \label{eom1}\\
        A_E'' - A_E'^2 + \frac{2A_E'}{z} + \frac{4}{9} \phi'^2 + \frac{\beta}{6} e^{\phi} (2\chi_u'^2 + \chi_s'^2) &= 0, \\
         A_t'' + A_t'\left(A_E' - \frac{1}{z} + \frac{\phi'\partial_\phi \omega}{\omega}\right) &= 0, \\         
         \phi'' + \left(\frac{f'}{f} + 3A_E' - \frac{3}{z}\right) \phi' - \frac{3 e^{2A_E} \partial_\phi V_E}{8 z^2 f} \notag\\
         - \frac{3\beta}{16} e^{\phi} (2\chi_u'^2 + \chi_s'^2) + \frac{3z^2 e^{-2A_E} A_t'^2 \partial_\phi \omega}{4f} \notag\\
         - \frac{3\beta e^{2A_E} \partial_\phi (e^\phi V)}{8 z^2 f} &= 0, \\
        \chi_u'' + \left(\frac{f'}{f} + 3A_E' - \frac{3}{z} + \phi'\right) \chi_u' - \frac{e^{2A_E} \partial_{\chi_u} V}{2 z^2 f} &= 0, \\
         \chi_s'' + \left(\frac{f'}{f} + 3A_E' - \frac{3}{z} + \phi'\right) \chi_s' - \frac{e^{2A_E} \partial_{\chi_s} V}{z^2 f} &= 0.  \label{eom5}
 \end{align}

 The system is solved under the following boundary conditions:
\begin{align}
     f(0) = 1, \quad f(z_h) &= 0,     \quad\phi'(0) = p_1,\\
     A_t(0) = \mu_B, \quad A_t(z_h) &= 0, \quad
     \chi_{u,s}'(0) = m_{u,s},
 \end{align}
where \(\mu_B\) is the baryon chemical potential, and \(m_{u,s}\) denote the light and strange quark masses. This construction provides the foundation for a dynamically consistent and data-driven holographic model capable of describing the thermodynamics and chiral dynamics of 2+1-flavor QCD.


\textit{Observables and computational scheme}. To explore the QCD phase structure, we analyze the thermodynamics and phase transitions of the 2+1-flavor EMDf system. The temperature \(T\) and entropy density \(s\) are computed from the black hole horizon data as
\begin{equation}
    T =\frac{|f'(z_h)|}{4\pi}, \qquad s =\frac{2\pi e^{3 A_E(z_h)}}{\kappa_N^2 z_h^3}.
\end{equation}
At zero chemical potential, the pressure \(p\) and energy density \(\varepsilon\) are obtained via standard thermodynamic identities:
\begin{equation}
    dp = s\,dT, \qquad \varepsilon = -p + sT.
\end{equation}

The full expressions for the chiral condensates are obtained via holographic renormalization, using the renormalized on-shell action $S_r \equiv (S + S_\partial)_{\text{on-shell}}$, where $S_\partial$ includes the necessary boundary contributions, such as the Gibbons-Hawking term and appropriate counterterms introduced to cancel divergences at the asymptotic boundary. The thermal chiral condensates for light and strange quarks are then defined by the functional derivatives of the renormalized action \(S_r\) with respect to the corresponding quark masses:
\begin{align}
  \langle\bar{\psi} \psi\rangle_u^T &= \frac{\delta S_{r}}{\delta m_{u}} = \frac{\beta}{\kappa_N^2}\sigma_u + b_1,   \\ 
  \langle\bar{\psi} \psi\rangle_s^T &= \frac{\delta S_{r}}{\delta m_{s}} = \frac{\beta}{\kappa_N^2}\sigma_s + b_2,
\end{align}
where \(b_1\) and \(b_2\) are scheme-dependent constants determined by the holographic renormalization procedure, and \(\sigma_{u,s}\) are extracted from the UV asymptotics of the bulk scalar fields:
\begin{align}
    \chi_{u,s}(z\to 0) = m_{u,s} z + \cdots + \sigma_{u,s} z^3 + \cdots.
\end{align}

For direct comparison with lattice results, the thermal chiral condensates are normalized as:
\begin{align}
\text{hotQCD: } &\quad \frac{\langle \bar{u} u \rangle_T}{\langle 0 | \bar{u} u | 0 \rangle} = 1 - \frac{\hat{d} - \Delta_u^R(T)}{\hat{d} - \Delta_u^R(\infty)}, \label{eq:hotQCD_u} \\
&\quad \frac{\langle \bar{s} s \rangle_T}{\langle 0 | \bar{s} s | 0 \rangle} = 1 + \frac{\hat{d} - \Delta_s^R(T)}{2 m_s r_1^4 \langle 0 | \bar{s} s | 0 \rangle}, \label{eq:hotQCD_s} \\
\text{W-B: } &\quad \Delta_{l,s} = \frac{\langle \bar{\psi} \psi \rangle_u^T - \frac{m_u}{m_s} \langle \bar{\psi} \psi \rangle_s^T}{\langle \bar{\psi} \psi \rangle_u^0 - \frac{m_u}{m_s} \langle \bar{\psi} \psi \rangle_s^0}, \label{eq:WB}
\end{align}
where $\Delta_q^R = \hat{d} + 2m_s r_1^4 \left[\langle\bar{\psi} \psi\rangle_q^T - \langle\bar{\psi} \psi\rangle_q^0\right]$ with $\hat{d}=0.0232244$ and $r_1=0.3106$ fm. The vacuum quark condensate $\langle0| \bar{s}s|0\rangle=-(0.307\text{GeV})^3$.
This normalization aligns with the lattice QCD prescriptions \cite{Gubler:2018ctz, Borsanyi:2010bp}, ensuring consistency between the holographic observables and lattice definitions.

Model parameters are constrained by matching the EoS and chiral observables to lattice QCD results at zero chemical potential. This ensures that the holographic description captures key nonperturbative features of QCD thermodynamics. Parameters relevant to $\mu_B=0$ include $\gamma$, $\gamma_1$, $b_4$, $\lambda$, $d_1$, $d_2$, and $G_5$, along with boundary inputs $p_1$, $m_u$, and $m_s$. The extension to finite density involves three additional parameters, $c_1$, $c_2$, and $c_3$.

Given the larger parameter space and the need to fit multiple observables, we implement a machine learning-based optimization strategy integrated with spectral methods. The optimization strategy consists of two main stages. In the first stage, a deep neural network is trained on lattice QCD data using the Adam optimizer to map temperature to key observables—the entropy density and the normalized chiral condensate. The network architecture includes three hidden layers with ReLU activation and is optimized using a mean squared error (MSE) loss function. After training, the network provides predictions that support accurate fitting and reliable extrapolation. In the second stage, the outputs of the pretrained neural network are combined with the holographic predictions. A new loss function, defined as the MSE between the neural network outputs and the holographic calculations, is then minimized via gradient descent using the Adam optimizer to fit and calibrate the model parameters with high precision.

\begin{figure}[htbp]
    \centering
\includegraphics[width=0.9\linewidth]{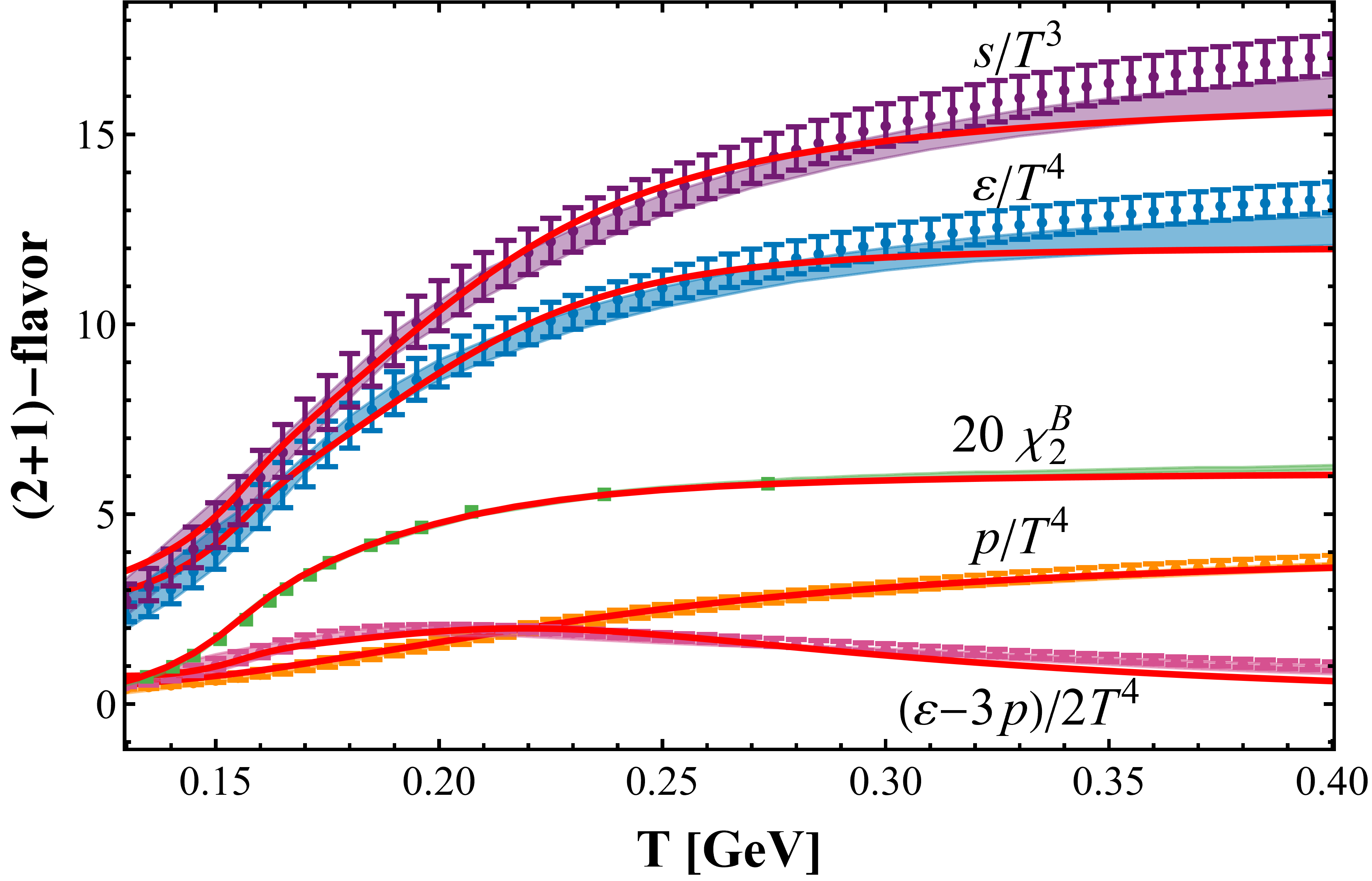}
    \caption{Thermodynamic observables from the EMDf model (red solid lines) compared with lattice QCD results (points with error bars and error bands represent hotQCD \cite{HotQCD:2014kol,Bazavov:2017dus} and W-B \cite{Borsanyi:2013bia,Borsanyi:2021sxv} lattice simulations respectively).}
    \label{fig:1}
\end{figure}

\begin{figure}[htbp]
    \centering
    \includegraphics[width=0.88\linewidth]{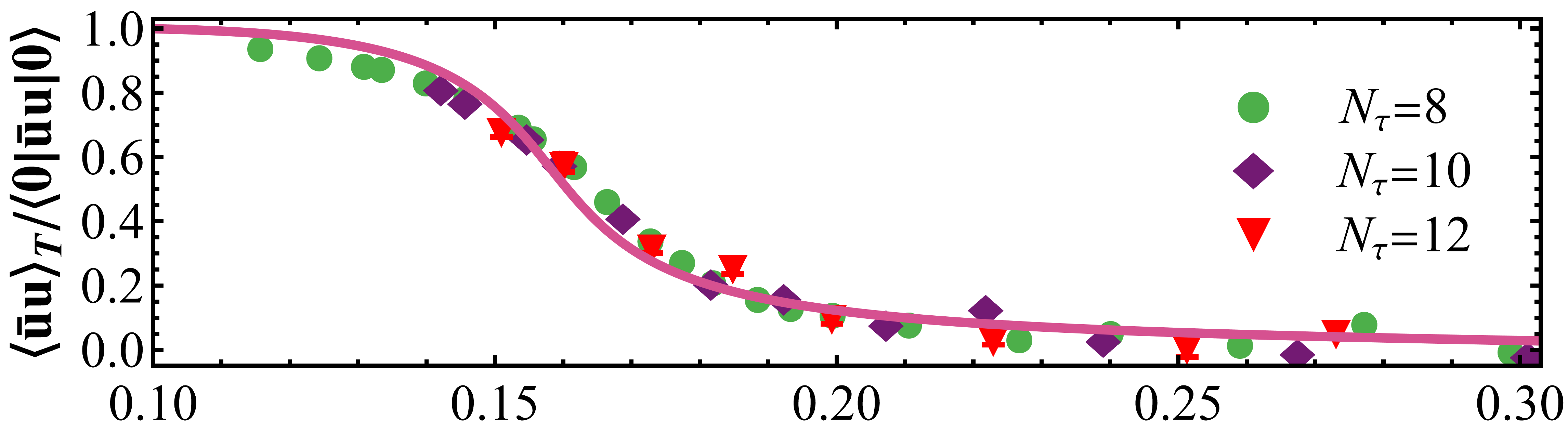}
    \vspace{0cm}
    \hspace{-0.295cm}
    \includegraphics[width=0.9025\linewidth]{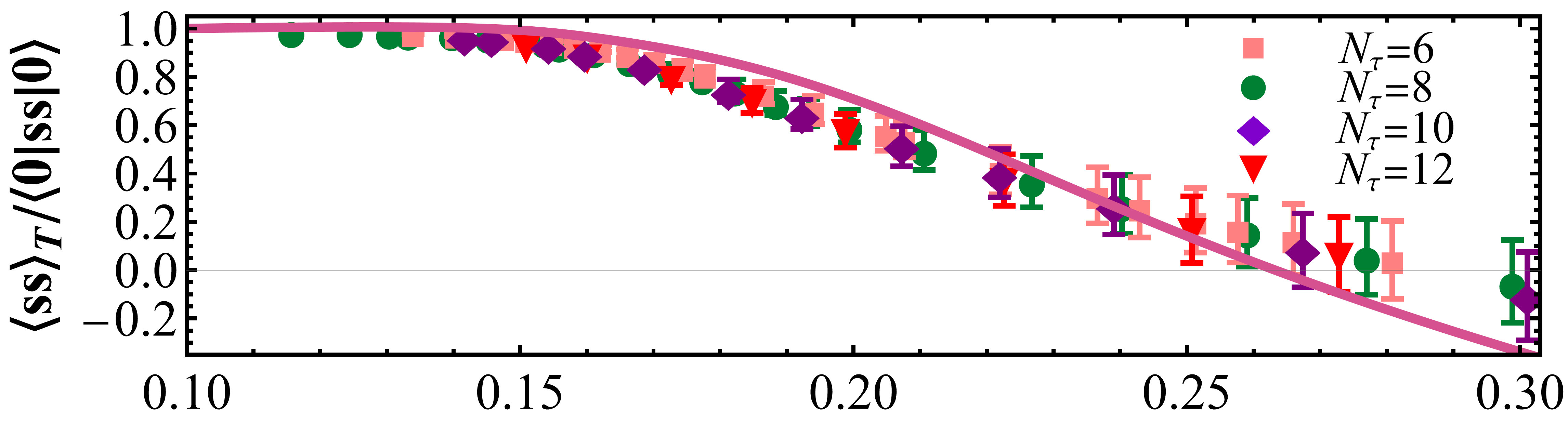}
    \vspace{0cm}
    \hspace{-0.15cm}
    \includegraphics[width=0.88 \linewidth]{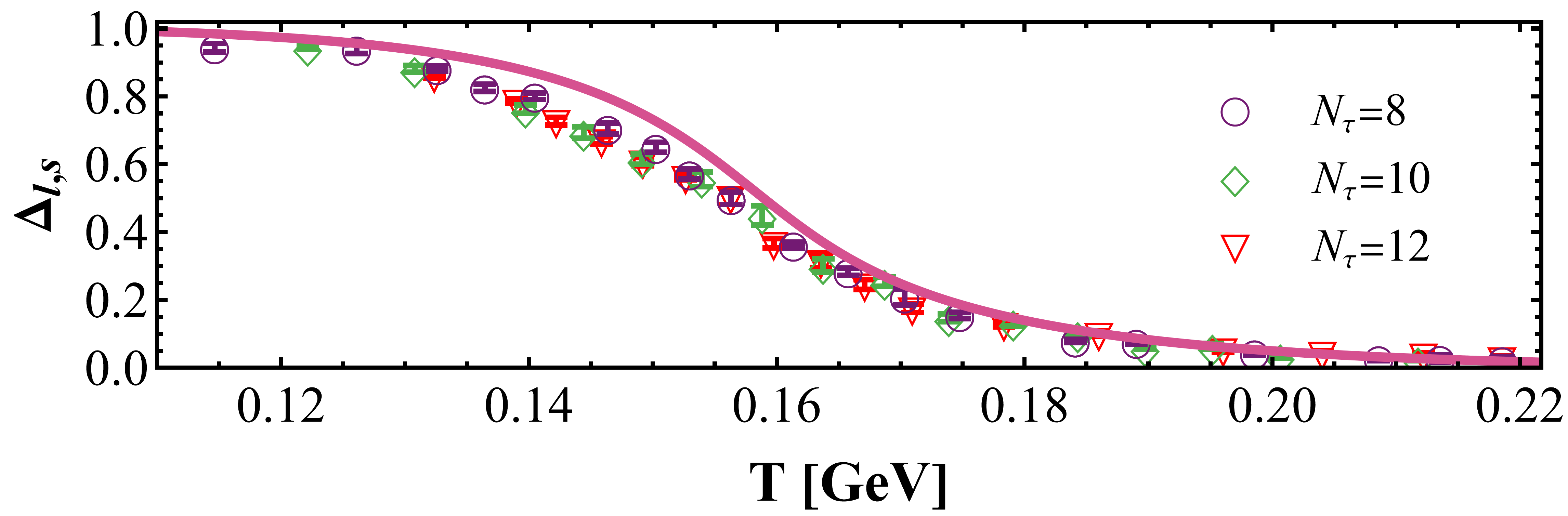}
    \caption{Comparison of chiral transitions for light (top) and strange (middle) quarks with hotQCD results \cite{HotQCD:2014kol,Gubler:2018ctz}, as well as the subtracted chiral condensate (bottom) compared with W-B results \cite{Borsanyi:2010bp}.}
    \label{fig:2}
\end{figure}

\textit{Thermodynamics and chiral transitions at $\mu_B =0$}. We examine the thermodynamics of the 2+1-flavor EMDf system at zero chemical potential. Since the model parameters are calibrated directly against lattice QCD data, the parameter-fitting procedure itself fixes the physical normalization, so that the holographic thermodynamic quantities can be compared with lattice observables without introducing additional scaling factors. The light and strange quark masses that best reproduce lattice QCD results are found to be $m_{u,d}^{\mathrm{phy}} = 3.5$ MeV and $m_s^{\mathrm{phy}} = 139$ MeV, respectively. The relevant model parameters at $\mu_B = 0$ are listed in Table~\ref{table:1}. The resulting thermodynamic observables—including the pressure $p$, energy density $\varepsilon$, entropy density $s$, trace anomaly $\varepsilon - 3p$, and baryon number susceptibility $\chi_2^B$---are shown in Fig.~\ref{fig:1}, while chiral transitions for light and strange quarks are displayed in Fig.~\ref{fig:2}. The model results exhibit excellent agreement with 2+1-flavor lattice QCD data~\cite{Borsanyi:2013bia,HotQCD:2014kol,Borsanyi:2021sxv,Borsanyi:2010bp,Gubler:2018ctz,Bazavov:2017dus}.

\begin{table}[htbp]
  \centering
  \begin{tabular}{c c c c c c c c}
    \hline\hline
    $\gamma$ & $\gamma_1$ & $b_4$ & $\lambda$ & $G_5$ & $d_1$ & $d_2$ & $p_1$ \\
    \hline
     1.7 & 0.737 & 0.12 & 2.4 & 0.48 & -0.86 & -0.115 & 0.365 \\
    \hline\hline
  \end{tabular}
  \caption{Fitted parameters of the EMDf model at $\mu_B = 0$. The parameter $p_1$ is given in units of GeV.}
  \label{table:1}
\end{table}

The coupling between the flavor fields and gravitational background, while complicating the numerical calibration, proves essential for capturing the nontrivial interplay between thermodynamic and chiral observables. The chiral crossover temperature, which is typically defined by the peak of $\left|d\langle\bar{\psi}\psi\rangle_u/dT\right|$ in holographic QCD \cite{Cao:2020ryx,Chen:2024jet,Bartz:2016ufc}, is found to be $T_\chi \simeq 158.2$ MeV,
in close agreement with the lattice estimate $T_\chi^{\mathrm{lat}} = (156.5 \pm 1.5)$ MeV~\cite{HotQCD:2018pds}. The minimum of the squared speed of sound $c_s^2$ occurs at $T_c \simeq 155.5$ MeV, also consistent with the lattice values \(154 \pm 9\) MeV~\cite{HotQCD:2014kol}.

The deconfinement transition temperature, estimated from the inflection point of the renormalized Polyakov loop, is $T_d \simeq 228.9$ MeV. Although this exceeds typical lattice values ($\sim 170$-200 MeV)~\cite{Bazavov:2009zn,Bazavov:2016uvm}, such a shift is expected due to the Polyakov loop being an approximate order parameter in full QCD. In the pure Yang-Mills limit, the deconfinement temperature is higher ($T_d^{\mathrm{gauge}} \simeq 285$ MeV)~\cite{Borsanyi:2022xml,Giusti:2025fxu}. The observed hierarchy $T_\chi < T_d$ reflects the partial decoupling between chiral symmetry restoration and deconfinement, consistent with lattice results. This separation hints at the possibility of an intermediate regime with mixed quark-gluon degrees of freedom~\cite{Cohen:2023hbq,Fujimoto:2025sxx}, a feature naturally realized in the present holographic framework.

\begin{figure}[htbp]
    \centering   \includegraphics[width=0.7\linewidth]{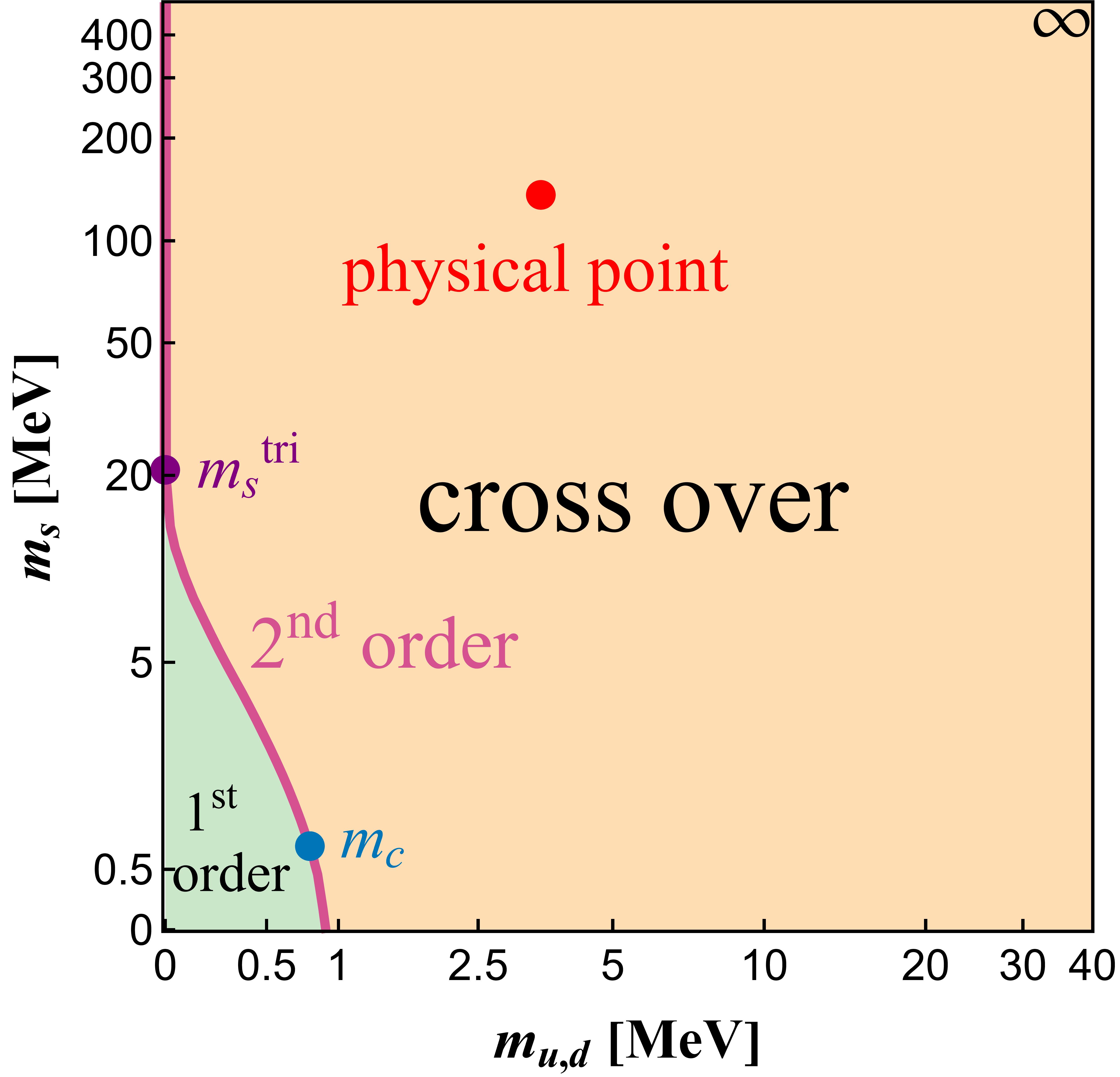}
    \caption{Quark mass phase diagram predicted by the 2+1-flavor EMDf model, showing the crossover and first-order regions separated by a second-order line.}
    \label{fig:3}
\end{figure}

\textit{Mapping the Columbia plot}. The dependence of QCD phase transitions on quark masses is encoded in the Columbia plot, a cornerstone of lattice QCD studies~\cite{Ding:2015ona}. Within our 2+1-flavor EMDf framework, we systematically explore this mass dependence by varying the light and strange quark masses. As shown in Fig.~\ref{fig:3}, the resulting phase diagram exhibits a second-order critical line (solid purple) that separates a first-order region from a crossover region. The physical point, located at $(m_{u,d}^{\mathrm{phy}}, m_s^{\mathrm{phy}}) = (3.5, 139)$ MeV, lies well within the crossover domain.

The second-order line intersects the $m_{u,d} = 0$ axis at $m_s^{\mathrm{tri}} = 21.2$ MeV, identifying the tri-critical point significantly below the physical strange quark mass. In the flavor-symmetric limit ($m_{u,d} = m_s$), we find a critical mass $m_c = 0.785$ MeV, below which the transition becomes first-order. This result is compatible with lattice estimates, which locate the boundary at $m_{u,d} = m_s \lesssim m_s^{\mathrm{phy}}/270$~\cite{Endrodi:2007gc,Ding:2011du}.

Our EMDf model thus reproduces the key features of the Columbia plot across a broad range of quark masses, including the location of critical line and the nature of the phase transitions. The success of this mapping underscores the essential role of dynamical flavor-gravity coupling in capturing the full non-perturbative QCD phase structure within a bottom-up holographic approach.

\begin{figure}[htbp]
    \centering
    \includegraphics[width=0.8\linewidth]{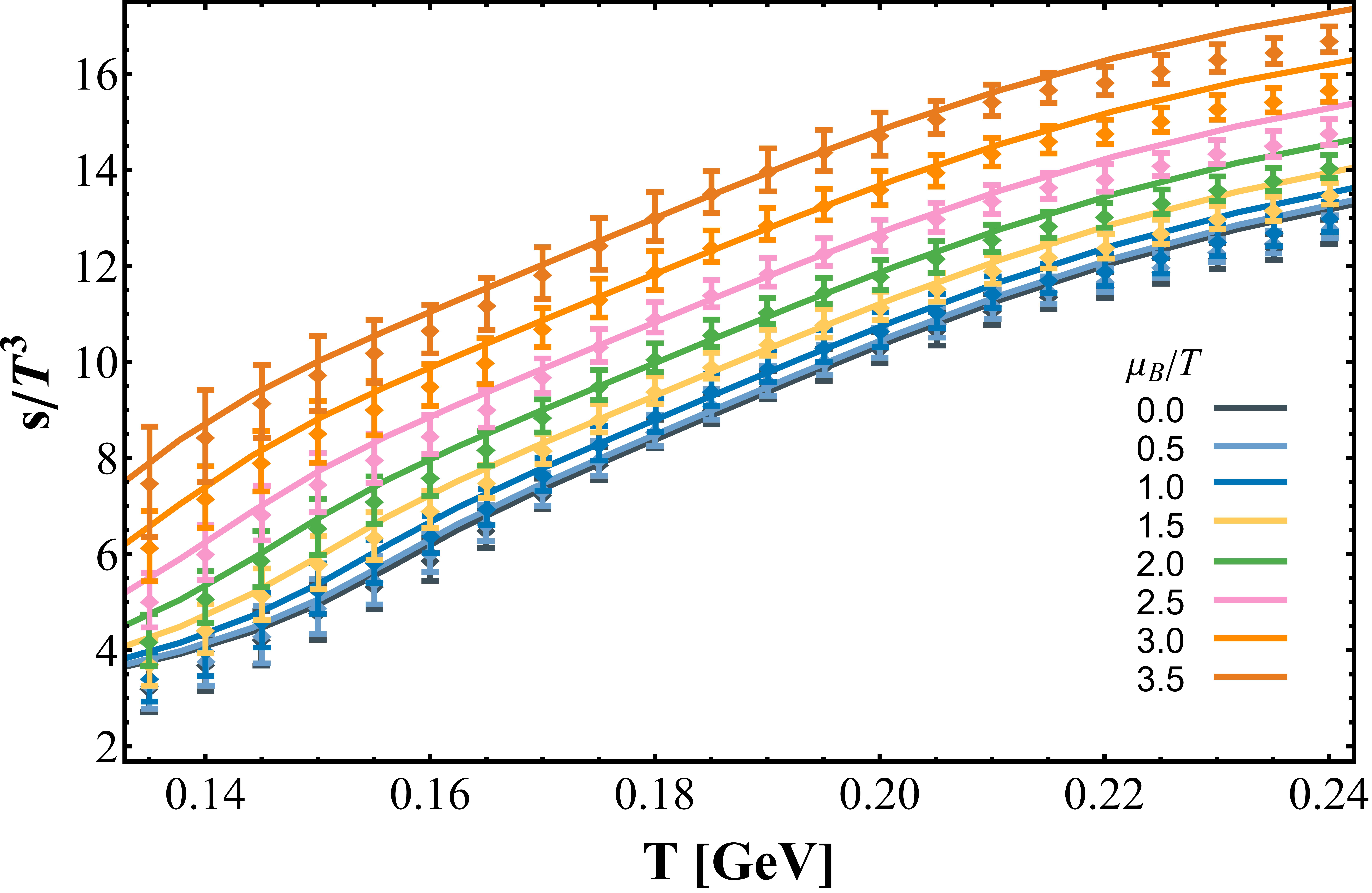}
    \includegraphics[width=0.8\linewidth]{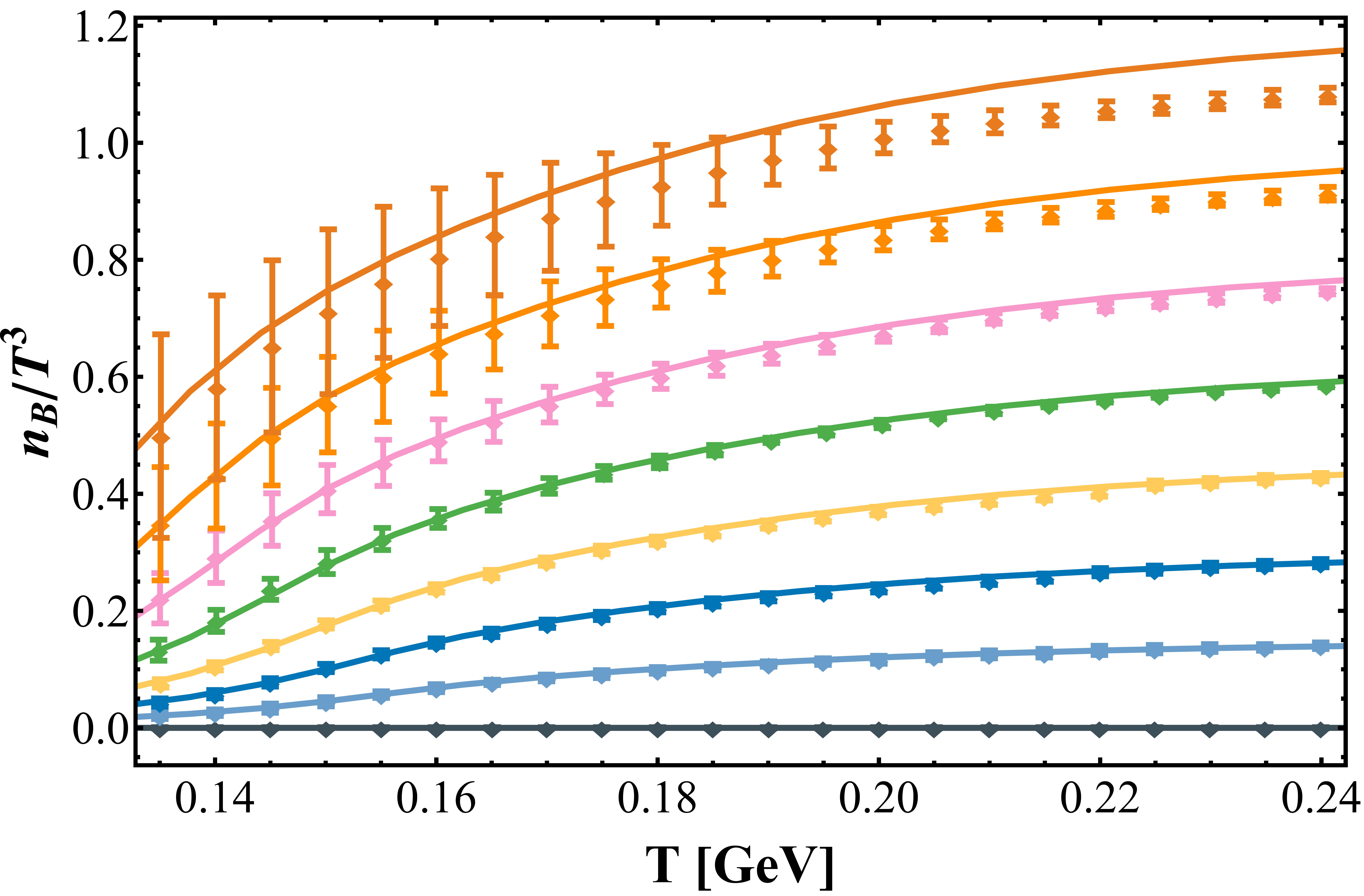}
    \caption{Entropy and baryon number density at various \(\mu_B/T\). Points with error bars denote lattice QCD data~\cite{Borsanyi:2021sxv}.}
    \label{fig:4}
\end{figure}

\textit{QCD phase diagram at finite density}. To investigate the QCD phase structure at finite chemical potential, we determine the charge sector parameters \(c_1 = 0.0975\), \(c_2 = 0.8\), and \(c_3 = 0.28\) by fitting the second-order baryon number susceptibility \(\chi_2^B = \partial(n_B/T^3)/\partial(\mu_B/T)\) to lattice QCD data (see Fig.~\ref{fig:1}). The baryon number density \(n_B\) is extracted from the near-boundary expansion of the bulk gauge field \(A_t\),
\begin{align}\label{eom55}
  A_t(z\to 0) = \mu_B -  \frac{\kappa_N^2 n_B}{4 \omega(0)} z^2 + \cdots.
\end{align}
As shown in Fig.~\ref{fig:4}, the model effectively reproduces the thermodynamic observables including the entropy and baryon number density across a wide range of \(\mu_B/T\), in good agreement with lattice QCD results~\cite{Borsanyi:2021sxv}.

\begin{figure}[htbp]
    \centering
    \includegraphics[width=0.9\linewidth]{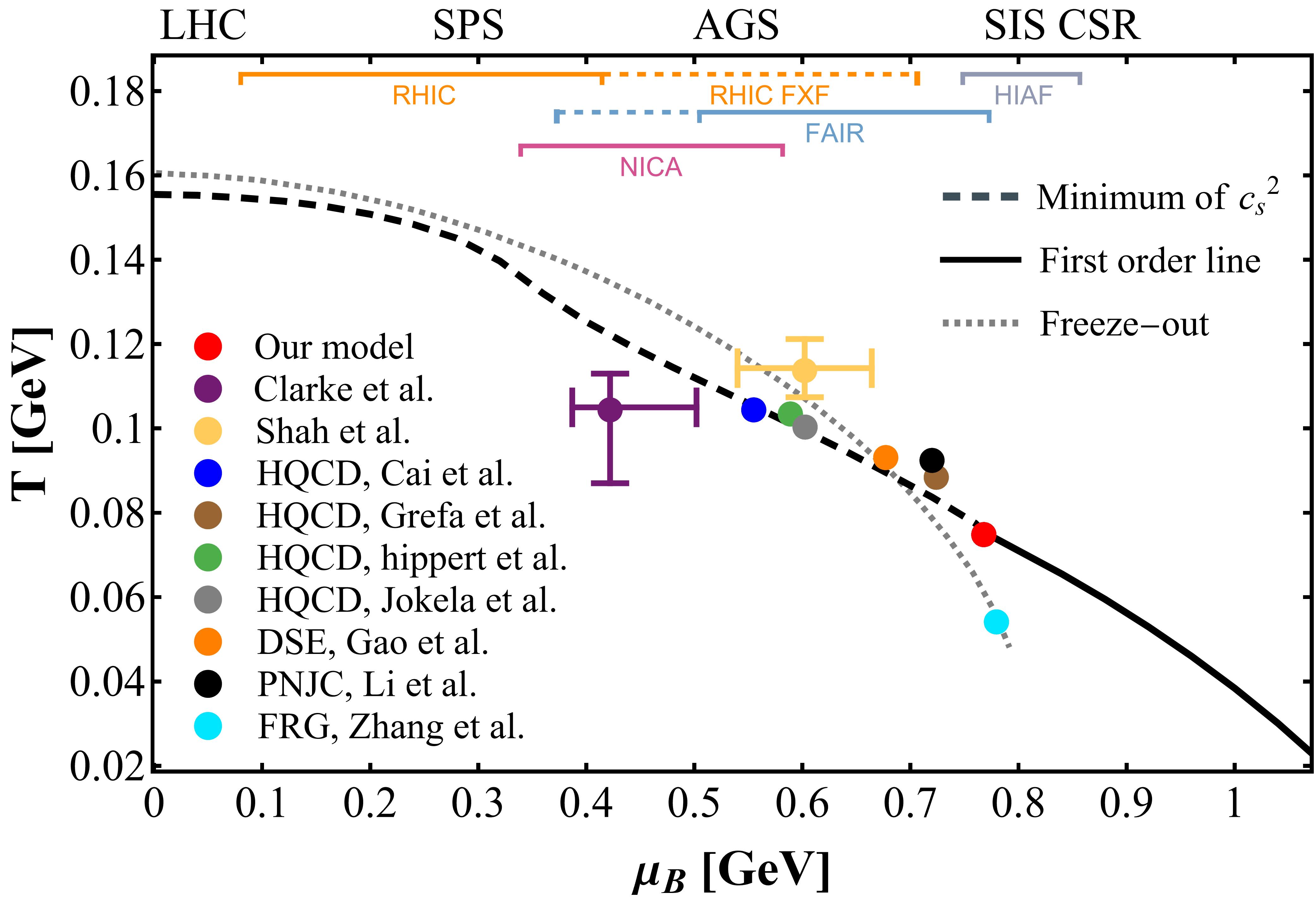}
    \caption{
QCD phase diagram from the EMDf model. Dashed and solid lines represent the crossover and first-order transition lines, respectively, with the CEP indicated by a red dot. Prior theoretical estimates of the CEP are included~\cite{Cai:2022omk,Gao:2020qsj,Li:2018ygx,Zhang:2017icm,Clarke:2024ugt,Shah:2024img,Grefa:2021qvt,Hippert:2023bel,Jokela:2024xgz}. The freeze-out line is depicted by gray dotted lines~\cite{Lysenko:2024hqp}, while top bars highlight the chemical freeze-out regions relevant to heavy-ion collision experiments~\cite{Luo:2020pef}.
}

    \label{fig:5}
\end{figure}

We map the QCD phase diagram in the \((T, \mu_B)\) plane, as displayed in Fig.~\ref{fig:5}. In the crossover region, where no definitive critical temperature exists, we define the transition temperature at the minimum of $c_s^2$. For first-order phase transition, it is determined by the intersection point of the free-energy line. The CEP is found at
\begin{align}
T_C = 75.4\ \mathrm{MeV}, \quad \mu_C = 768\ \mathrm{MeV},
\end{align}
which is compared with earlier holographic and QCD-based results~\cite{Cai:2022omk,Gao:2020qsj,Li:2018ygx,Zhang:2017icm,Clarke:2024ugt,Shah:2024img,Grefa:2021qvt,Hippert:2023bel,Jokela:2024xgz}. We also display the experimentally extracted freeze-out line \cite{Lysenko:2024hqp}, with the predicted CEP located above it. The experimental coverage shown in Fig.~\ref{fig:5} suggests that the predicted CEP lies within reach of upcoming heavy-ion collision programs \cite{Luo:2020pef}. 

\textit{Conclusion and discussion}. We have shown that the EMDf system, incorporating dynamical light and strange quark sectors, provides a unified holographic description of the QCD phase structure across varying quark masses and finite baryon density. Employing a machine-learning-assisted spectral method, the model achieves remarkable agreement with 2+1-flavor lattice QCD for both thermodynamic and chiral observables, marking the first instance of a bottom-up holographic QCD framework that consistently and quantitatively captures both deconfinement and chiral phase transitions.

The model captures the essential features of the QCD transition at zero density. The pseudo-critical temperatures for the chiral crossover (\(T_\chi \simeq 158\) MeV) and the minimum of the speed of sound (\(T_c \simeq 155\) MeV) agree with lattice values, while the Polyakov loop inflection point at \(T_d \simeq 229\) MeV confirms the hierarchy \(T_\chi < T_d\). The separation between chiral and deconfinement transitions suggests a semi-QGP regime, naturally realized in the holographic setup.

In the quark-mass plane, the model reproduces the Columbia plot structure, including the second-order critical line and the tri-critical point at \(m_s^{\mathrm{tri}} \simeq 21\) MeV. In the flavor-symmetric limit, the first-order transition terminates at a critical mass \(m_c \simeq 0.785\) MeV, consistent with lattice estimates. At finite density, the model yields a $(T,\mu_B)$ phase diagram with a crossover-to-first-order transition and a CEP at \(T_C = 75.4\) MeV and \(\mu_C = 768\) MeV, within reach of future heavy-ion experiments.

These findings demonstrate that the EMDf system dynamically realizes the 2+1-flavor QCD phase structure with quantitative accuracy. They establish the viability of combining nonperturbative holography with data-driven methods to probe strongly coupled matter, and pave the way for future explorations of QCD dynamics under high-density and non-equilibrium conditions.

\textit{Acknowledgments}. We are grateful to Danning Li, Xun Chen, Hong-An Zeng, and Ling-Jun Guo for their valuable discussions. This work is supported by Hunan Provincial Natural Science Foundation of China (Grants No. 2023JJ30115 and No. 2024JJ3004) and the YueLuShan Center for Industrial Innovation (2024YCII0117), and also supported by the National Key Research and Development Program of China under Grant No. 2020YFC2201501 and the National Science Foundation of China (NSFC) under Grants No.~12347103 and No.~11821505.

\bibliographystyle{apsrev4-1}
\bibliography{ref1}

\end{document}